\documentclass[preprint,number,sort&compress,twocolumn,3p]{elsarticle}
\pdfoutput=1
\usepackage{rotating}
\usepackage{graphicx}
\usepackage{array}
\usepackage{amsmath}
\usepackage{color}
\usepackage{amssymb}
\usepackage{slashed}
\usepackage[pdftex,hyperfootnotes=false,pdfpagelabels]{hyperref}

\def\bm#1{\boldsymbol{#1}}

\begin{document}

\title{On \texorpdfstring{Mono-$W$}{Mono-W} Signatures in \texorpdfstring{Spin-$1$}{Spin-1} Simplified Models}

\tnotetext[t1]{This article is registered under preprint number: CERN-TH-2016-047, DESY 16-042, UCI-HEP-TR-2016-05}

\author[add1,add2]{Ulrich Haisch}
\ead{Ulrich.Haisch@physics.ox.ac.uk}
\author[add3]{Felix Kahlhoefer}
\ead{felix.kahlhoefer@desy.de}
\author[add4]{and Tim M.P. Tait}
\ead{ttait@uci.edu}

\address[add1]{Rudolf Peierls Centre for Theoretical Physics, University of Oxford, \\ 1 Keble Road, Oxford OX1 3NP, United Kingdom}
\address[add2]{CERN, Theory Division, CH-1211 Geneva 23, Switzerland}
\address[add3]{DESY, Notkestra\ss e 85, D-22607 Hamburg, Germany}
\address[add4]{Department of Physics and Astronomy, University of California, Irvine, \\ California 92697, USA}

\begin{abstract}
The potential sensitivity to isospin-breaking effects makes LHC searches for mono-$W$ signatures promising probes of the coupling structure between the Standard Model and  dark matter.  It has been shown, however, that the strong sensitivity of the mono-$W$ channel to the relative magnitude and sign of the up-type and down-type quark couplings to dark matter is an artefact of unitarity violation. We provide three different solutions to this mono-$W$ problem in the context of spin-$1$ simplified models and briefly discuss the impact that our findings have on the prospects of mono-$W$ searches at future LHC runs.
\end{abstract}


\maketitle

\section{Introduction}

At least two reasons make the process $p p \rightarrow W + E_{T, \rm miss}$ interesting in the context of dark matter (DM) searches at the LHC. First, for leptonically decaying $W$ bosons, this process yields a distinct signature that can be searched for in dedicated analyses, which suffer from significantly less background than for instance mono-jet searches. Second, hadronic $W$-boson decays  will instead lead to a ${\rm jets} + E_{T, \rm miss}$ final state and  therefore render an electroweak (EW) contribution to the mono-jet channel.\footnote{Note that in spite of the name, mono-jet searches do not usually veto events with two jets. Indeed, even higher jet multiplicities are included in the latest LHC searches.} 

These so-called mono-$W$ searches have received significant interest because they are potentially sensitive to the sign between the up-type and down-type quark couplings to~DM~\cite{Bai:2012xg}. The reason is that two different diagrams contribute to this process and therefore interference effects can be important. Indeed, LHC Run I analyses~\cite{Aad:2013oja, CMS:2013iea, Khachatryan:2014tva} based on an effective field theory (EFT) approach to parameterise the interactions of DM \cite{Beltran:2010ww,Goodman:2010ku,Bai:2010hh} found a striking difference in the predicted fiducial cross sections between same-sign~(SS) and opposite-sign~(OS) couplings. In the OS case, leading to constructive interference between up-type and down-type quark contributions, the mono-$W$ results in fact turn out to set the strongest limits on the suppression scale of the unknown mediating interaction, surpassing the EFT limits that arise from mono-jet searches.\footnote{The ATLAS collaboration has very recently also searched for a mono-$W$ signal in $13 \, {\rm TeV}$ data \cite{ATLAS:13TeV},  upgrading their latest $8  \, {\rm TeV}$ analysis \cite{ATLAS:2014wra}.} In direct detection experiments, on the other hand, OS couplings lead to destructive interference and correspondingly smaller event rates, making the LHC a particularly promising probe for this scenario.

\begin{figure*}[htb]
\begin{center}
\includegraphics[width=0.7\textwidth]{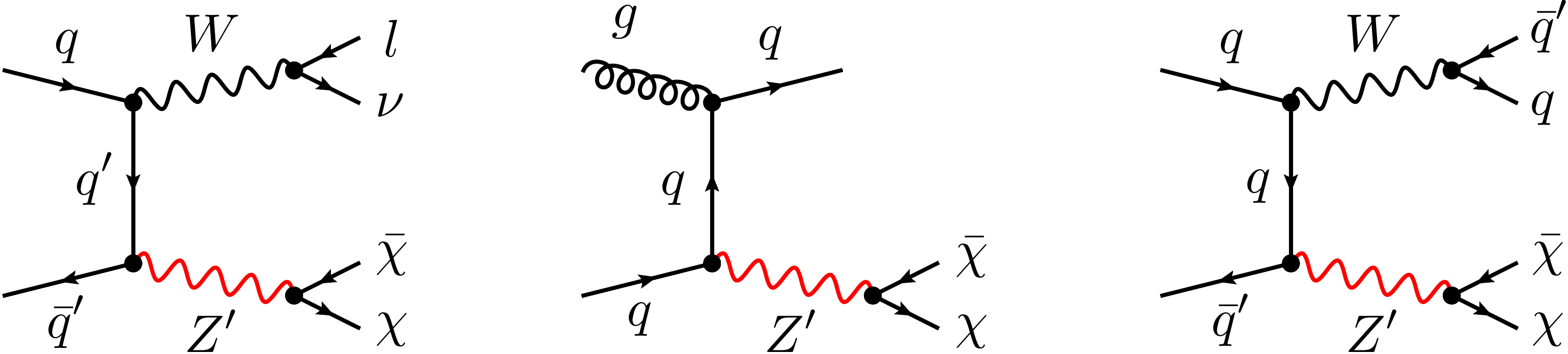} 
\end{center}
\vspace{0mm}
\caption{\label{fig:diagrams} Prototypes of Feynman diagrams that lead to a mono-$W$  (left) and a mono-jet signal~(middle and right). While the graph in the middle exemplifies the QCD contribution to this process, the diagram on the right-hand side represents an EW correction.}
\end{figure*}

Subsequently, it has however been pointed out in \cite{Bell:2015sza} that  EFT interpretations of mono-$W$ signals  have issues with unitarity violation, if the  effective higher-dimensional operators  break the $SU(2)_L$ invariance of the Standard Model (SM). This is for example the case when considering pure vector  DM-quark operators with OS Wilson coefficients for up-type and down-type quarks, as done in~\cite{Aad:2013oja, CMS:2013iea, Khachatryan:2014tva}. The enhancements of the mono-$W$ cross section found in these analyses is thus spurious, because it is due to the emission of longitudinal $W$ bosons. The EFT study~\cite{Bell:2015sza}  has recently been extended in \cite{Bell:2015rdw} to the case of simplified models with $t$-channel exchange of a new coloured scalar. In particular, it has been shown that if these theories are formulated before EW symmetry breaking~(EWSB) in an $SU(2)_L$ gauge-invariant way, effects associated with isospin breaking are generically small. This leads to the conclusion that isospin-violating DM-quark couplings are not expected to increase the sensitivity of  mono-$W$ searches in $t$-channel simplified models. It was also pointed out in \cite{Bell:2015rdw} that a similar solution is expected to be present in $s$-channel simplified models.

The goal of this note is to fully develop the reasoning presented in~\cite{Bell:2015sza,Bell:2015rdw} for $s$-channel DM models with vector or axial-vector mediator exchange. In Section~\ref{sec:problem} we demonstrate that a naive calculation of mono-$W$ signatures in spin-$1$  simplified models leads to unphysical predictions, in particular the violation of unitarity at large energies. We point out that this issue also affects mono-jet searches once EW corrections are included in the signal prediction. In Section~\ref{sec:solution1}, we propose a simple solution to this problem based on imposing certain requirements on the coupling structure of the simplified model. A different solution is derived in Section~\ref{sec:solution2} by considering mono-$W$ signatures in the SM, where the potentially dangerous terms are cancelled via interference with a diagram involving a triple gauge boson interaction. We discuss how this solution can be extended to simplified DM models. In~Section~\ref{sec:mixing} it is then shown that this second solution is indeed the one naturally incorporated in $Z^\prime$ models that obtain non-universal couplings to quarks from mixing with the SM gauge bosons. In~Section~\ref{sec:solution3}, we briefly discuss how unitarity violation can be parameterised by non-renormalisable interactions. The main findings of this note are summarised in  Section~\ref{sec:summary}, while the impact of EW corrections in an existing mono-jet search are studied in Appendix~\ref{app:EWmonoj}.

\section{The \texorpdfstring{mono-$\bm{W}$}{mono-W} problem in simplified models}
\label{sec:problem}

In order to illustrate the issues that can appear in the calculation of mono-$W$ signals in simplified DM models, we consider a spin-1 mediator $Z^\prime_\mu$ with mass $M_{Z^\prime}$ and a Dirac DM particle $\chi$ with mass $m_\text{DM}$. We write the relevant interaction terms in the Lagrangian as
\begin{equation}\label{eq:L_VA}
\begin{split}
\mathcal{L} =  & - Z^\prime_\mu \, \bar{\chi} \left[ g_\text{DM}^V \gamma^\mu + g_\text{DM}^A \gamma^\mu \gamma_5 \right] \chi \\[1mm] & - \sum_{f = q,l,\nu} Z^\prime_\mu \, \bar{f} \left[ g_{f}^V \gamma^\mu + g_f^A \gamma^\mu \gamma_5 \right] f \, .
\end{split}
\end{equation}
In the context of such a simplified model, the mono-$W$ signature arises at the LHC from the process $p p \rightarrow W + Z^\prime$, followed by the decay of the $Z^\prime$ into DM particles. A possible Feynman diagram leading to a $l + E_{T, \rm miss}$ final state is shown on the left-hand side of Figure~\ref{fig:diagrams}. Notice that the interactions (\ref{eq:L_VA}) also give rise to mono-jet signatures. There are in fact both QCD and EW corrections to jets~+~$E_{\rm T,miss}$, which are exemplified by the middle and right diagram in Figure~\ref{fig:diagrams}, respectively.

\begin{figure}[t]
\begin{center}
\includegraphics[width=0.75\columnwidth]{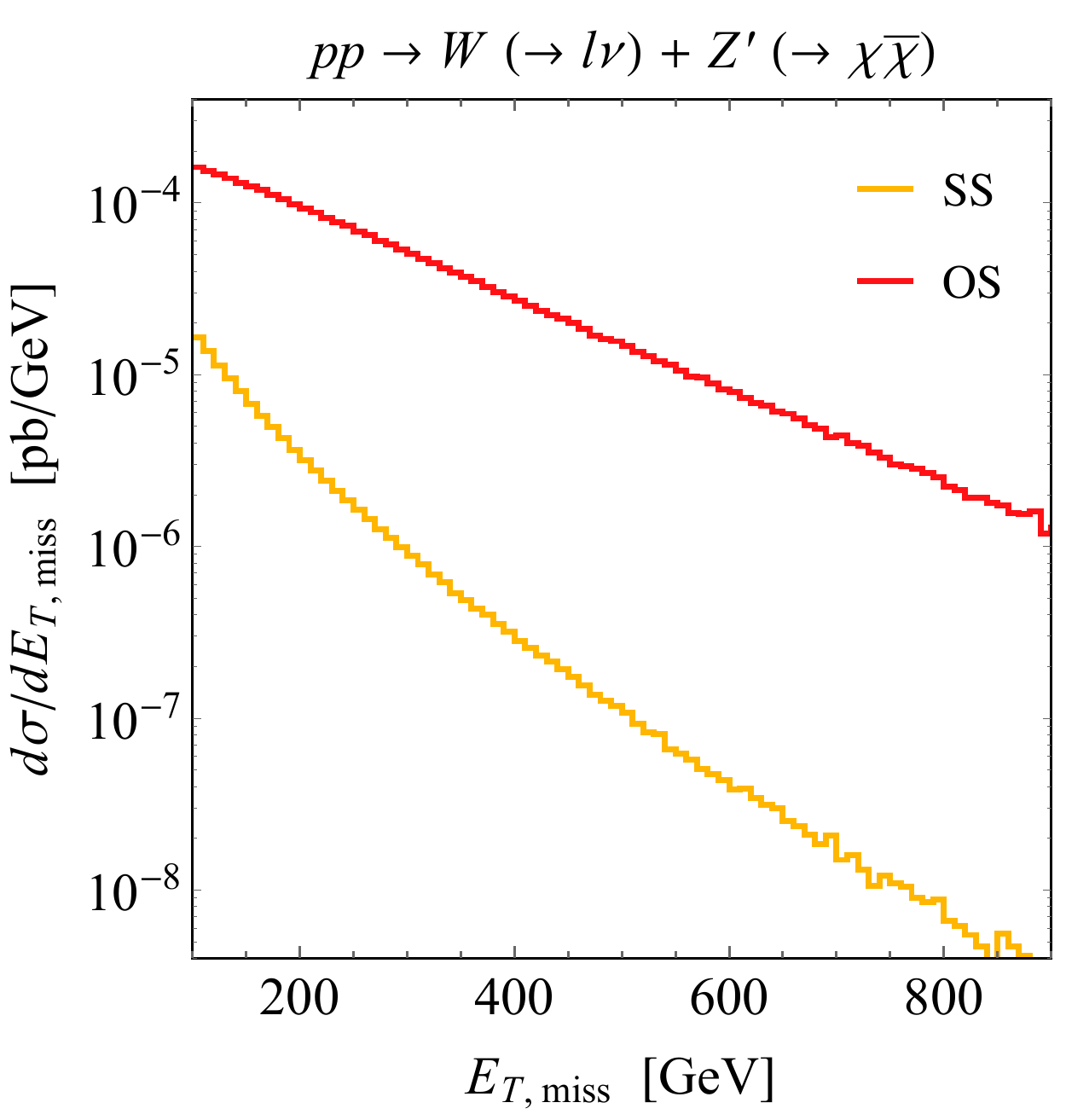} \qquad
\includegraphics[width=0.75\columnwidth]{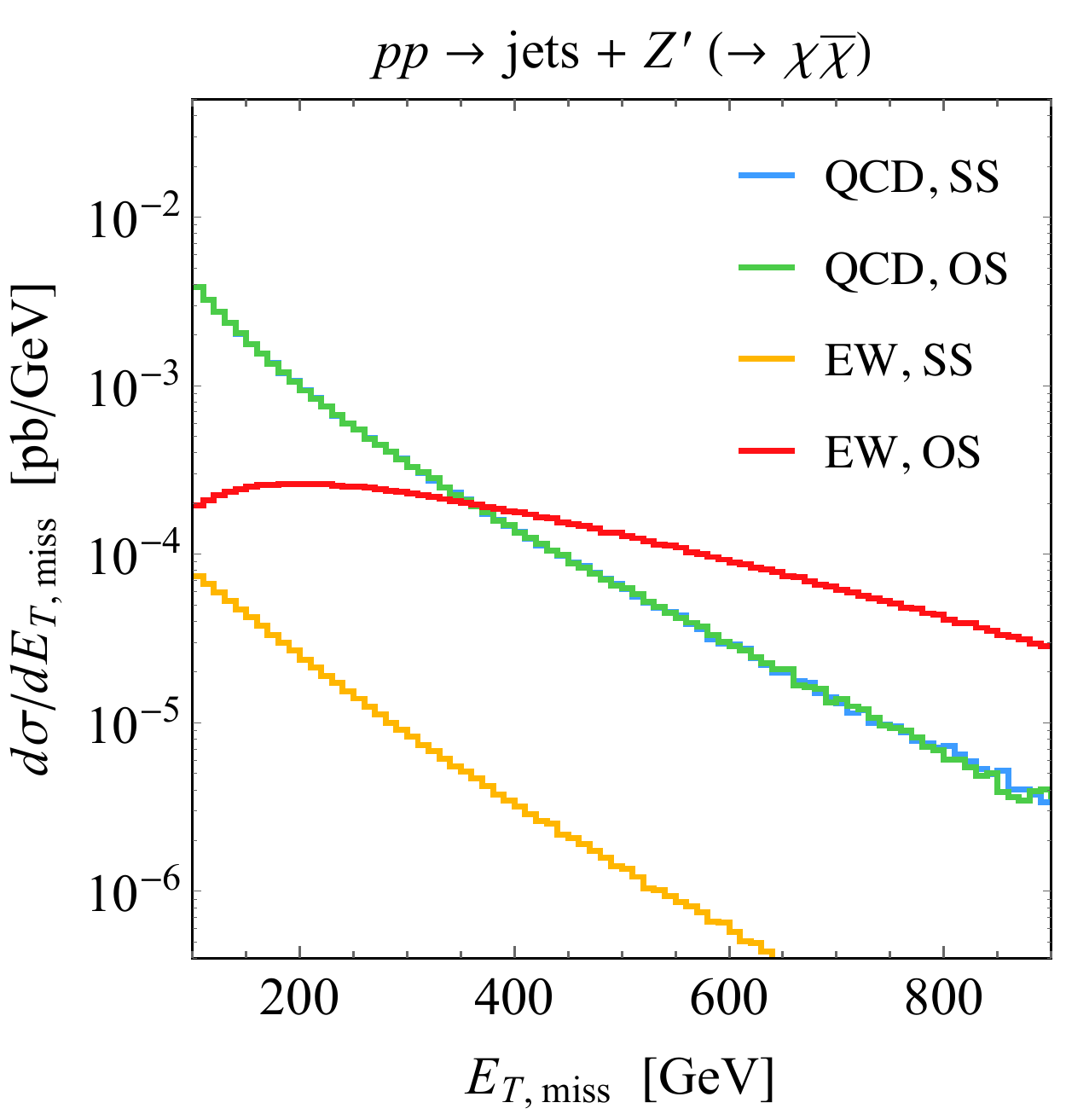}
\end{center}
\vspace{-4mm}
\caption{\label{fig:METspectra} Example of a $E_{T, \rm miss}$ spectrum of a mono-$W$ (top panel) and mono-jet (bottom panel) signal arising in the spin-$1$ simplified model. See text for additional explanations.}
\end{figure}

The fundamental problem discussed in this paper is illustrated in Figure~\ref{fig:METspectra}, which shows the predicted $E_{T, \rm miss}$ spectrum for leptonically (left panel) and hadronically (right panel) decaying $W$ produced in association with an invisibly decaying $Z^\prime$. Our event generation has been performed at leading order with {\tt MadGraph5\_aMC\@NLO}~\cite{Alwall:2014hca} starting from the implementation of the spin-$1$ simplified model presented in~\cite{Backovic:2015soa} and utilises {\tt NNPDF2.3} parton distribution functions (PDFs)~\cite{Ball:2012cx}. We consider $pp$ collisions at a center-of-mass energy of $\sqrt{s} = 8 \, {\rm TeV}$, employ $M_{Z^\prime} = 1 \, {\rm TeV}$, $\Gamma_{Z^\prime} = 56.5 \, {\rm GeV}$, $m_{\rm DM} = 10 \, {\rm GeV}$, $g_{\rm DM}^V = 1$ and  set all axial-vector and leptonic couplings to zero. The  yellow curves in both panels correspond to  SS couplings $g^V_u = g^V_d = 0.25$, while the red curves represent the OS coupling choice $g^V_u = -g^V_d = -0.25$. It is evident from the left plot in Figure~\ref{fig:METspectra} that the mono-$W$  predictions for the two coupling choices do not simply differ by a  rescaling factor, but that the predicted differential cross sections are fundamentally different. Most notably the $E_{T,\rm miss}$ spectrum  in the OS case is significantly harder than that for the SS choice. 

The suspicious high-energy behaviour of the $pp \to W + Z^\prime$ amplitudes in the OS case becomes particularly obvious in the right panel of Figure~\ref{fig:METspectra}, where we compare the $E_{T,\rm miss}$ spectrum for hadronically decaying $W$ to the $E_{T,\rm miss}$ spectrum for a conventional mono-jet analysis, where the jets arise from QCD interactions (see middle and right diagram in Figure~\ref{fig:diagrams}). In the SS case, one observes that the $E_{T,\rm miss}$ spectra for both processes are similar, but that the overall cross section for the QCD process~(blue curve) is significantly larger than that of the EW process~(yellow curve), the latter amounting to a relative correction of a few percent only. In the OS case, on the other hand, we find the EW contribution~(red curve) to the differential cross section of  $pp \to {\rm jets}  + E_{T,\rm miss}$ to be harder and to dominate over the QCD process~(green curve) at sufficiently large $E_{T, \rm miss}$ values. This finding turns out to be essentially unaffected by  the parton shower, hadronisation corrections and detector effects.  For spin-$1$ simplified model realisations~(\ref{eq:L_VA}) with OS couplings, the EW channel $pp \to W \hspace{0.25mm} (\to q \bar q^\prime) + Z^\prime  \hspace{0.25mm} (\to \chi \bar \chi)$ and not the QCD process $pp \to {\rm jets} + Z^\prime  \hspace{0.25mm} (\to \chi \bar \chi)$ is therefore expected to give rise to the majority of events in  the  high-$E_{T,\rm miss}$ signal regions~(SRs) of LHC Run~I mono-jet searches. That this expectation is indeed correct is shown  in Appendix~\ref{app:EWmonoj}.

The paradoxical observation that an EW contribution appears to produce harder mono-jet events than a QCD process casts doubt on the validity of the spin-$1$ simplified model introduced in  (\ref{eq:L_VA}). Indeed, we will show below that the harder  $E_{T,\rm miss}$ spectrum in the OS case is due to contributions that grow with energy and therefore potentially violate unitarity. The aim of this note is to accurately describe the nature of this problem and propose a number of ways of solving it by appropriately modifying or restricting the spin-$1$ simplified model under consideration.\footnote{Note that, since the interactions (\ref{eq:L_VA}) of the $Z^\prime$ boson explicitly break the EW symmetry, the corresponding Ward identities are no longer satisfied by $W$ bosons in the final state. As a result, the Goldstone boson equivalence theorem does not hold,~i.e.~one does not obtain the same result at high energies when replacing $W_L$ fields by Goldstone bosons. In other words, since the gauge symmetry is broken, unitary gauge and Feynman gauge are not equivalent and one cannot simply remove the mono-$W$ problem by calculating cross sections in Feynman gauge.}

\section{Unitarity violation and coupling structures}
\label{sec:solution1}

It is well-known that the production of longitudinal gauge bosons (such as $Z^\prime_L$) from fermions can potentially violate unitarity at large energies (see for instance~\cite{Babu:2011sd}). For example, it was shown in~\cite{Kahlhoefer:2015bea} that for the spin-$1$ simplified model considered in (\ref{eq:L_VA})  the process $\chi + \bar \chi \rightarrow Z^\prime_L + Z^\prime_L$ violates unitarity at large energies for non-zero axial-vector couplings, unless a second dark Higgs is added to the theory. Since the amplitude is proportional to the fermion mass, the corresponding process with light quarks only violates unitarity at very large energies and can therefore be disregarded.

The process $u + \bar{d} \rightarrow W^+_L + Z^\prime_L$, however, can lead to much lower scales of unitarity violation. Following~\cite{Chanowitz:1978mv,Kahlhoefer:2015bea}, we calculate the helicity matrix elements for the case of a left-handed up-quark and a right-handed anti-down quark (both of which are charged under $SU(2)_L$). Since the initial state has a total spin of 1, the process cannot proceed at $s$-wave~($\mathcal{M}^0 = 0$). For the $J = 1$ partial wave (or $p$-wave), we find in the limit $s \rightarrow \infty$
\begin{equation} \label{eq:M1}
\mathcal{M}^1 =  \frac{g \hspace{0.25mm} V_{ud} \hspace{0.25mm} s}{96 \pi \hspace{0.25mm} M_W \hspace{0.25mm}  M_{Z^\prime}} \left(g^A_u-g^A_d-g^V_u+g^V_d\right) \, ,
\end{equation}
with $g$ the $SU(2)_L$ gauge coupling and $M_W$ the mass of the $W$ boson. The helicity matrix element is found to grow proportional to $s$ at large energies. For
\begin{equation}
\label{eq:unitarityviolation}
 s > \frac{48\pi \hspace{0.25mm} M_W \hspace{0.25mm}  M_{Z^\prime}}{g \, |V_{ud}| \left | g^A_u-g^A_d-g^V_u+g^V_d\right |} \,,
\end{equation}
it will therefore violate the requirement for perturbative unitarity, $|\mathcal{M}^J |< 1/2$.\footnote{The matrix element corresponding to $u + \bar{d}  \rightarrow W^+_L + Z^\prime_T$ with $Z^\prime_T$ denoting a transversal $Z^\prime$ boson scales as $\sqrt{s}/M_W \left(g^A_u-g^A_d-g^V_u+g^V_d\right)$ in the high-energy limit. In consequence, the resulting mono-$W$ cross section does not grow with $s$, but becomes a constant in the $s \to \infty$ limit.  Compared to the purely longitudinal contribution (\ref{eq:M1}), transversal modes thus provide only a subleading source of unitarity violation.}

In the SS case  with vanishing axial-vector couplings, one finds for the combination of couplings appearing in (\ref{eq:unitarityviolation}) that $g^A_u-g^A_d-g^V_u+g^V_d = 0$ and hence the potentially unitarity-violating term vanishes identically. For OS couplings, on the other hand, one has $g^A_u-g^A_d-g^V_u+g^V_d = - 2 g^V_u \neq 0$, if the axial-vector couplings are zero. The resulting mono-$W$ and mono-jet cross sections then grow proportional to $s$, which explains both the large enhancement of the total fiducial cross section and the harder $E_{T, \rm miss}$ spectrum observed in Section~\ref{sec:problem} for the OS case. 

The most obvious way to avoid this problem is to simply impose the requirement
\begin{equation} \label{eq:cond1}
g^A_u-g^A_d-g^V_u+g^V_d = 0 \,,
\end{equation}
on the coupling structure of the spin-$1$ simplified model (\ref{eq:L_VA}).  In order to understand this requirement intuitively, let us rewrite the left-handed and right-handed couplings in terms of  vector and axial-vector couplings. For the sign conventions adopted in (\ref{eq:L_VA}), one obtains 
\begin{equation}
  g_q^L = g_q^V - g_q^A \,,\qquad
 g_q^R = g_q^V +  g_q^A \,.
\end{equation}
In terms of left-handed and right-handed couplings, the requirement (\ref{eq:cond1}) hence simply becomes $g_u^L = g_d^L$,~i.e.~left-handed up-type and down-type quarks should couple in the same way to the $Z^\prime$ mediator. 

Notice that the latter requirement would be automatically fulfilled, if one wrote the quark couplings to the spin-$1$ mediator in such a way so as to preserve the EW symmetry,
\begin{align} 
{\cal L}_{Z^\prime q \bar q} = & - \sum_{u,d} \, Z^\prime_\mu \left[ g_u \hspace{0.5mm}  \bar{u}_R  \hspace{0.25mm}  \gamma^\mu \hspace{0.1mm}  u_R + g_d  \hspace{0.5mm}  \bar{d}_R \hspace{0.25mm}  \gamma^\mu \hspace{0.1mm}  d_R  \right]  \nonumber \\
& -\sum_{Q} \, Z^\prime_\mu  \, g_Q \hspace{0.5mm} \bar{Q}_L  \hspace{0.25mm}  \gamma^\mu \hspace{0.1mm}  Q_L\,,
\end{align}
where $Q_L = (u_L, d_L)^T$ is the left-handed quark doublet and $u_R$ and $d_R$ are the right-handed quark singlets.  Each quark flavour has couplings described by the three parameters $g_Q$, $g_u$ and $g_d$, and one can arrange for any desired relative sign between either the vector or axial-vector couplings, but  cannot choose them independently from one another.\footnote{We emphasise that there are additional considerations that may further restrict the possible choices of $g_Q$, $g_u$ and $g_d$. For example, to be consistent with minimal flavour violation, the same couplings should be chosen for all three quark families~\cite{Abdallah:2015ter}. Moreover, models with non-zero axial-vector couplings (i.e.\ $g_u \neq g_Q$ or $g_d \neq g_Q$) require additional structure in the Higgs sector, which may lead to strong constraints from Higgs measurements or EW precision observables~\cite{Kahlhoefer:2015bea}.} 

It is however conceivable that the coupling structure of the $Z^\prime$ boson is modified by EWSB and that the spin-1 mediator ends up coupling differently to left-handed up-type and down-type quarks. In this case, the unitarity issue of amplitudes like $u + \bar d \to W^+ + Z^\prime$  needs to be solved in a different fashion.

\section{Interference with additional diagrams}
\label{sec:solution2}

In the previous section, we have shown that unitarity violation in the process $p p \rightarrow W+ Z^\prime$ can be avoided if we require that the $Z^\prime$ couples with equal strength to left-handed up-type and down-type quarks. We can compare this requirement to the couplings of the SM $Z$ boson, whose interactions with SM quarks can be written as
\begin{equation}
 \mathcal{L}_{Z q \bar q} = - \sum_q Z_\mu \, \bar{q}\left[ g_{Z,q}^V \hspace{0,25mm}  \gamma^\mu  + g_{Z,q}^A  \hspace{0,25mm}   \gamma^\mu \gamma_5  \right] q \, ,
\end{equation}
with 
\begin{equation}
\begin{split}
g_{Z,q}^V & =  \frac{g}{2 \hspace{0.25mm}  c_w} \left(T_{3,q} - 2 Q_q \hspace{0,25mm} s_w^2\right) \,, \\[1mm]
g_{Z,q}^A & =  -\frac{g}{2  \hspace{0.25mm}  c_w}  \hspace{0.5mm} T_{3,q} \,,
\end{split}
\end{equation}
where $s_w$ and $c_w$ are the sine and cosine of the weak mixing angle, $Q_q$ denotes the electromagnetic quark charge in units of the elementary charge $e$ and the weak isospin is given by $T_{3,u} = 1/2$ and $T_{3,d} = -1/2$. It follows that the difference $\Delta g_Z$ of the left-handed couplings of the $Z$ boson to up-type and down-type quarks reads
\begin{align}
\Delta g_Z & = g_{Z,u}^L - g_{Z,d}^L = g_{Z,u}^V - g_{Z,u}^A - g_{Z,d}^V + g_{Z,d}^A \nonumber \\ 
& = g \hspace{0.25mm} c_w \neq 0 \,.
\end{align}
In other words, the requirement (\ref{eq:cond1}) is not satisfied by the $Z$ boson within the SM and as a result, one might expect that the process $p p \rightarrow W + Z$ also shows a unitarity-violating behaviour at high energies.

\begin{figure*}
\begin{center}
\includegraphics[width=0.6\textwidth]{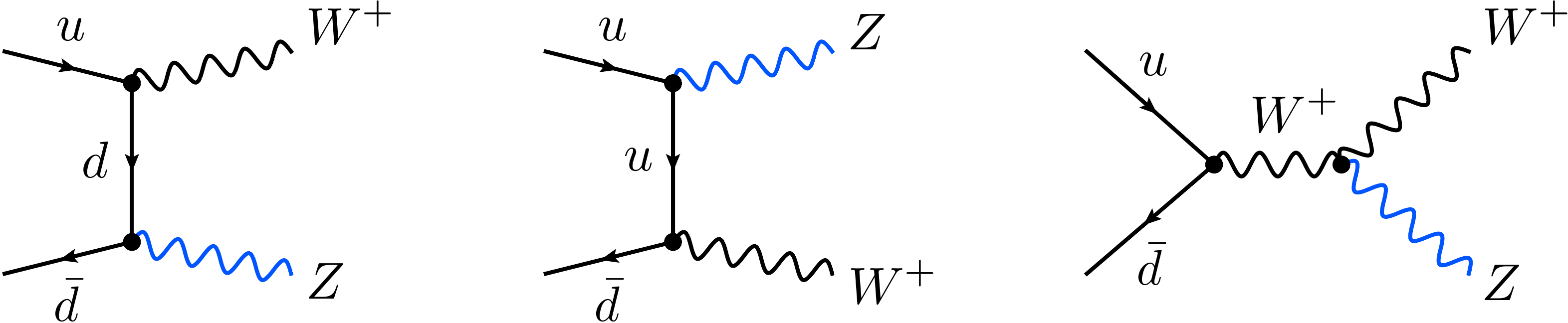} 
\end{center}
\vspace{0mm}
\caption{\label{fig:wzdiagrams} The three tree-level diagrams that contribute to $u + \bar d \to W^+ + Z$ in the SM. The left and middle graph involves only couplings between quarks and gauge bosons, while the right diagram contains a triple gauge boson vertex.}
\end{figure*}

Indeed, when considering only the two $t$-channel diagrams shown on the left-hand side and in the middle of Figure~\ref{fig:wzdiagrams} the resulting  amplitude grows with energy. In the SM, however, a third diagram is present, which involves a $s$-channel $W$ boson radiating off a $Z$ boson. The corresponding graph is depicted on the right-hand side in the latter figure. It involves a triple gauge boson vertex of the form 
\begin{equation}
\label{eq:LWWZ}
\mathcal{L}_{WWZ} =  i  g_{WWZ}  \hspace{0.25mm}  {\cal T}_{WWZ} \,,
\end{equation} 
with $g_{WWZ} = g \hspace{0.25mm} c_ w = \Delta g_Z$ and
\begin{equation}  \label{eq:TWWV}
\begin{split}
 {\cal T}_{WWV} & = \Bigl[ \left (\partial_\mu W_\nu^+ - \partial_\nu W_\mu^+ \right ) W^{\mu -} V^{\nu} \\[1mm]
  & \phantom{xxx} - \left (\partial_\mu W_\nu^- - \partial_\nu W_\mu^- \right ) W^{\mu +} V^{\nu}  \\[1mm]
  & \phantom{xxx} + \frac{1}{2} \left (\partial_\mu V_\nu - \partial_\nu V_\mu \right ) \\[1mm]
  & \phantom{xxxx} \times \left (W^{\mu +} W^{\nu -} - W^{\mu -} W^{\nu +} \right ) \Bigr] \,.
\end{split}  
\end{equation}
Combining the three different contributions, one finds that in the high-energy limit  the square of the $u + \bar d \to W^+ + Z$ matrix element is given by
\begin{equation} \label{eq:WZSM}
|{\cal M}|^2 = \frac{3 g^4 c_w^4 |V_{ud}|^2}{32 M_W^2} \left (d_1 + d_2 - 2 d_3 \right )^2 \, s^2 \hspace{0.25mm} \sin^2 \theta   \,,
\end{equation}
where $V_{ud}$ denotes the relevant element of the Cabibbo-Kobayashi-Maskawa matrix, $\theta$~is the scattering angle in the center-of-mass frame and $d_1$, $d_2$ and $d_3$ label the contributions from the three different graphs. From the result (\ref{eq:WZSM}) it is readily seen that while the individual diagrams all diverge,  their sum remains finite at large energies, since $d_1 + d_2 - 2 d_3 =0$ for $d_1=d_2=d_3 =1$. Notice that the cancellation of unitarity-violating terms among the diagrams of Figure~\ref{fig:wzdiagrams} is not at all accidental, but a direct consequence of the  local $SU(2)_L$ gauge invariance of the SM. 

The above observation motivates a second solution to the mono-$W$ problem.\footnote{This solution was also briefly discussed in~\cite{Bell:2015rdw}.} If the difference $\Delta g$ between the left-handed couplings of the $Z^\prime$ boson to up-type and down-type quarks takes the non-zero value 
\begin{equation} \label{eq:Deltag}
 \Delta g = g_u^L - g_d^L = g_u^V - g_u^A - g_d^V + g_d^A \,,
\end{equation}
it is still possible to avoid unitarity-violating processes by adding additional gauge boson interactions of the form 
\begin{align}
\label{eq:solution2}
\Delta \mathcal{L} = i  g_{WWZ^\prime} \hspace{0.5mm}  {\cal T}_{WWZ^\prime} \, ,
\end{align}
with $g_{WWZ^\prime} = \Delta g$ to the Lagrangian (\ref{eq:L_VA}) of the spin-$1$ simplified model. Here ${\cal T}_{WWZ^\prime}$ denotes the $WWZ^\prime$ analogue of the $WWV$ interactions introduced in (\ref{eq:TWWV}). 

\section{A \texorpdfstring{${\bm Z^\prime}$}{Z'} boson with mixing}
\label{sec:mixing}

In this section we discuss a class of models where the additional interactions (\ref{eq:solution2})  between the $W$ bosons and the $Z^\prime$ boson are automatically generated in such a way that they cancel the potentially unitarity-violating contributions to the mono-$W$ signal. The basic idea is that the mass eigenstate $Z^\prime_\mu$ arises from  an interaction eigenstate vector field~$X_\mu$ with the SM $U(1)_Y$ $B_\mu$ field and the neutral component $W^3_\mu$ of the $SU(2)_L$ weak fields through mixing 
\begin{align}
\label{eq:general-mixing}
\left(\begin{array}{c} B_\mu \\ W^3_\mu \\ X_\mu \end{array}\right)=
\left(\begin{array}{ccc}  N_{11} & N_{12} & N_{13}
\\ N_{21} & N_{22} & N_{23}
\\ N_{31} & N_{32} & N_{33}
\end{array}\right)
\left(\begin{array}{c} A_\mu \\ Z_\mu \\ Z_\mu^\prime \end{array}\right) \, .
\end{align}
Here $A_\mu$ and $Z_\mu$ denotes the physical photon and neutral massive gauge boson fields of the SM. At this point, we leave the entries  $N_{ij}$ of the above mixing matrix unspecified. Examples for how such a mixing matrix can be generated and a very brief discussion of the constraints on the $N_{ij}$ that arise from EW precision tests will be given below.

The couplings of the interaction eigenstate $X_\mu$ can be parameterised in the same way as in the spin-$1$ simplified model introduced in Section~\ref{sec:problem}. Including only DM and quark couplings, we write 
\begin{equation}\label{eq:L_X}
\begin{split}
\mathcal{L} =  & - X_\mu \, \bar{\chi} \left[ f_\text{DM}^V \gamma^\mu + f_\text{DM}^A \gamma^\mu \gamma_5 \right] \chi \\[1mm]
& - \sum_{q} X_\mu \, \bar{q} \left[ f_{q}^V \gamma^\mu + f_q^A \gamma^\mu \gamma_5 \right] q \, , 
\end{split}
\end{equation}
and assume that the direct couplings $f_q^V$ and $f_q^A$ satisfy the requirement (\ref{eq:cond1}),~i.e.~they fulfill $f_u^L - f_d^L = f^A_u-f^A_d-f^V_u+f^V_d = 0$. In the absence of mixing unitarity-violating contributions to the mono-$W$ process will hence be absent.

However, as we have discussed in Section~\ref{sec:solution2}, the SM $Z$ boson couples differently to left-handed up-type and down-type quarks and in the presence of mixing the $Z^\prime$ can inherit this difference. The induced couplings between the $Z'$ and SM fermions are discussed for example in~\cite{Frandsen:2012rk}.\footnote{Notice that the sign convention for the axial-vector couplings used in~\cite{Frandsen:2012rk} is opposite to the one adopted in the present work.} For the left-handed quark couplings, one obtains 
\begin{equation} \label{eq:Zpinduced}
\begin{split}
g_u^L & = f^L_u \hspace{0.25mm} N_{33} + \frac{g}{2} \hspace{0.25mm} N_{23} + \frac{e}{6 c_w} \hspace{0.25mm} N_{13} \, , \\[1mm] 
g_d^L & = f^L_d \hspace{0.25mm} N_{33} - \frac{g}{2} \hspace{0.25mm} N_{23} + \frac{e}{6 c_w} \hspace{0.25mm} N_{13} \, .
\end{split}
\end{equation}
Since by assumption  $f^L_u - f^L_d = 0$ and due to the SS of the terms proportional to $N_{13}$, the coupling difference (\ref{eq:Deltag})  thus takes the simple form 
\begin{equation}
\Delta g = g_u^L - g_d^L  = g \hspace{0.25mm} N_{23} \,.
\end{equation}

In addition to the  couplings (\ref{eq:Zpinduced}), the mixing matrix~$N$ entering (\ref{eq:general-mixing}) also induces couplings of the $Z^\prime$ to other SM particles~\cite{Chun:2010ve}. In particular, the SM $WWZ$ vertex (\ref{eq:LWWZ}) will lead to  $WWZ^\prime$ interactions. Since the triple gauge boson vertex arises from the non-abelian $SU(2)_L$, the induced coupling turns out to be proportional to 
\begin{equation}
g_{WWZ^\prime} = g \hspace{0.25mm} N_{23} \,,
\end{equation}
and of the form (\ref{eq:TWWV}). The resulting $WWZ^\prime$ interaction  is therefore precisely that given in~(\ref{eq:solution2}).  As discussed in Section~\ref{sec:solution2}, the unitarity issue of the $u + \bar d \to W^+ + Z^\prime$ channel is then solved by the interference of the two diagrams with $t$-channel quark exchange and the single graph involving a $s$-channel $W$ boson.

It is also important to realise that in the class of models discussed in this section, the mono-$W$ channel in addition receives a contribution from $p p \to W + Z \hspace{0.25mm} (\to \chi \bar \chi)$, since the SM $Z$-boson field~$Z_\mu$ inherits the direct DM couplings $f_{\rm DM}^V$ and $f_{\rm DM}^A$ from the interaction field $X_\mu$ through mixing. Like  (\ref{eq:WZSM}) this contribution however remains unitary at large energies as a result of the $SU(2)_L$ gauge invariance of the~SM. This shows that the mono-$W$ problem is indeed absent in $Z^\prime$ models with mixing.

Since we did not specify the origin of the gauge boson mixings $N_{ij}$, our conclusions apply to a range of different scenarios. For example, a non-zero entry $N_{23}$ can be generated via kinetic mixing~\cite{Holdom:1985ag}
\begin{equation}\label{eq:kin-mix}
\mathcal{L} _{F^\prime  \hspace{-0.25mm} B} =  -\frac{1}{2}  \hspace{0.25mm} \sin \epsilon \, F^\prime_{\mu\nu}  B^{\mu\nu} \; ,
\end{equation}
between the $U(1)'$ field strength $ F^\prime_{\mu\nu}$ and the SM hypercharge field strength $B_{\mu \nu}$.  In this case, one finds in the limit of small mixing angles $\epsilon \ll 1$ the result 
\begin{equation}
N_{23} = - c_w \hspace{0.25mm} s_w \, \epsilon \, \frac{M_Z^2}{M_Z^2 - M_{Z^\prime}^2} \, ,
\end{equation}
with $M_Z$ denoting the mass of the SM $Z$ boson. Another option is to consider so-called mass mixing between the $Z$ and the $Z^\prime$~\cite{Babu:1997st}, which is described by 
\begin{equation}
 \mathcal{L}_{Z^\prime \hspace{-0.25mm} Z} = \delta m^2 \, Z^\mu Z'_\mu \, .
\end{equation}
The $N_{23}$ entry in the gauge boson mixing matrix takes for mass mixing the form 
\begin{equation}
N_{23} = -  c_w \, \frac{\delta m^2}{M_Z^2 - M_{Z^\prime}^2} \, . 
\end{equation}
We note for ${Z^\prime}$-boson masses in the TeV range both $\epsilon$ and $\delta m^2/M_Z^2$ have to be very small (of the order of $10^{-2}$ or below) in order not to violate bounds from EW precision measurements (see~e.g.~\cite{Frandsen:2012rk}). Nevertheless, there are also other ways to generate gauge boson mixing, for example from left-right symmetric models~\cite{Shu:2016exh}.

\section{Parametrising unitarity violation with non-renormalisable interactions}
\label{sec:solution3}

In the previous section, we have seen that $Z^\prime$ models with mixing remain unitary at high energies, if they are $SU(2)_L$ symmetric  in the absence of mixing. While this treatment covers most of the simplest models for a spin-1 mediator, there may of course be more complicated models with additional heavy particles at a new-physics scale $\Lambda$.  In such a model,  the couplings of the mediator to the left-handed up-type and down-type quarks may be realised via higher-dimensional operators such as
\begin{equation} \label{eq:LZpQH}
\begin{split}
{\cal L}_{Z^\prime Q H} = - \sum_{u,d} Z^\prime_\mu \Bigl[ & \frac{1}{\Lambda_u^2}  \,  (\bar{Q}_L \tilde{H}) \hspace{0.25mm} \gamma^\mu (\tilde{H}^\dagger Q_L) \\[1mm] 
& + \frac{1}{\Lambda_d^2}  \, (\bar {Q}_L H) \hspace{0.25mm} \gamma^\mu (H^\dagger Q_L)
\Bigr] \,,
\end{split}
\end{equation}
which effectively decouple the left-handed up-type and down-type quark couplings.  Here $\Lambda_{u,d}$ are two different suppression scales and $\tilde H = i \sigma_2 H^\ast$ with  $H$ denoting the $SU(2)_L$ Higgs doublet. In such a case, the difference $\Delta g$ between the left-handed quark couplings can  be parameterised by~\cite{Bell:2015sza}
\begin{equation} \label{eq:EFT}
\Delta g = \frac{v^2}{\Lambda^2} \,,
\end{equation}
with $v \simeq 246 \, {\rm GeV}$ the Higgs vacuum expectation value.\footnote{In the specific case (\ref{eq:LZpQH}), the new-physics scale $\Lambda$ appearing in (\ref{eq:EFT})  is given by $\Lambda = (\left |\Lambda_u^{-2} - \Lambda_d^{-2} \right | / 2 )^{-1/2}$.} Since the details of the ultraviolet completion are not specified, there is no direct connection between the $Z^\prime$ coupling to SM quarks and the $Z^\prime$ coupling to $W$ bosons. 

As a result, the mono-$W$ problem may still be present in such models. Notice that this is not a theoretical inconsistency, because unitarity violation is a generic feature of EFTs once the partonic centre-of-mass energy $\sqrt{s}$ (or more generically the momentum transfer relevant for the process under consideration) approaches the new-physics scale $\Lambda$.\footnote{The effect of unitarity violation in EFT models for DM has been discussed in~\cite{Shoemaker:2011vi,Fox:2012ee,Endo:2014mja}.} The mono-$W$ problems thus does not preclude meaningful calculations within the EFT parameterisation (\ref{eq:EFT}), provided the scale $\Lambda$ is sufficiently large. In fact, one can translate the bound  (\ref{eq:unitarityviolation}) into a limit on the new-physics scale $\Lambda$ for a given partonic centre-of-mass energy:
\begin{equation}
 \Lambda > \left ( \frac{s \hspace{0.5mm} v}{24 \pi M_{Z^\prime}} \right)^{1/2} \,.
\end{equation}
This means for instance that if one wants to study a model with $M_{Z^\prime} = 1 \, {\rm TeV}$, one needs to impose $\Lambda > 0.75 \, {\rm TeV}$ in order to ensure that the model does not violate perturbative unitarity up to $\sqrt{s} = 13 \, {\rm TeV}$. This in turn limits the magnitude of the couplings in the OS case to  $|g^V_u| = |g^V_d| < 0.05$.

Of course, the $\sqrt{s} = 13 \, {\rm TeV}$ LHC does on average not probe partonic center-of-mass energies of $13 \, {\rm TeV}$, so a smaller new-physics scale $\Lambda$ and therefore larger OS couplings may in practice be acceptable. A proper treatment is likely to require a self-consistent truncation procedure, which discards all events with a momentum transfer larger than $\Lambda$ (as discussed for instance in~\cite{Busoni:2013lha,Busoni:2014sya}). For example, to consistently study the OS example considered in Section~\ref{sec:problem}, one should only consider kinematic configurations with $\sqrt{s} < 6.1 \, {\rm TeV}$ (for $M_{Z'} = 1 \, {\rm TeV}$).  Such a truncation is expected to lead to a notable reduction of the LHC sensitivity to mono-$W$ signatures. 

\section{Summary}
\label{sec:summary}

In the context of simplified models with spin-1 mediators, we have discussed the so-called mono-$W$ problem,~i.e.~the violation of perturbative unitarity in the process $p p \rightarrow W + E_{T,\rm miss}$, and proposed three different solutions. These are 
\begin{enumerate}
 \item \textbf{Restricted coupling structures}: The mono-$W$ problem is absent if the coupling structure of the simplified model is restricted in such a way that the spin-1 mediator couples equally to left-handed up-type  and down-type quarks $g_u^L  = g_d^L$. This requirement is automatically satisfied if the interaction between the $Z^\prime$ and the quarks are formulated in an $SU(2)_L$ invariant way. 
 \item \textbf{Additional interactions}: For $g_u^L \neq g_d^L$, the mono-$W$ problem can still be solved if an additional interaction between the spin-1 mediator and SM $W$ bosons is introduced, which is proportional to the difference between the left-handed couplings. The interference between $t$-channel quark exchange and $s$-channel $W$-boson diagrams then removes the potentially dangerous contributions.
 \item \textbf{Small couplings}: If no additional interactions between the $Z^\prime$ and $W$ bosons are present, the mono-$W$ process will violate perturbative unitarity at large energies. The scale of unitarity violation may however be larger than the energies probed in a given process, if the difference between the couplings to up-type  and down-type quark couplings is sufficiently small.
\end{enumerate}
We have argued that the first solution is realised in the simplest $Z^\prime$ models, whereas the second solution appears naturally in $Z^\prime$-boson scenarios with mixing. If a more general model is to be considered, the third solution allows for a general assessment of the validity of a chosen coupling structure in a given experimental environment. 

The arguments presented in our work strongly suggest that the sensitivity of mono-$W$ searches to the parameter space of spin-$1$ simplified models cannot exceed that of the mono-jet channel. This conclusion seems to hold irrespectively of how the unitarity problem in the $pp \to W + E_{T,\rm miss}$ process is tamed. The same verdict has been previously reached in~\cite{Bell:2015sza,Bell:2015rdw} for the EFT case and $t$-channel simplified DM models with coloured scalar exchange.  It has also been argued in~\cite{Bell:2015rdw} that the mono-$W$ problem is not present in $s$-channel models if they are formulated in a gauge-invariant way, and our note explicitly proves this conjecture.
 
\section*{Acknowledgments}

We thank the members of the LHC DM working group for discussions that triggered this project and Nicole~Bell for carefully reading a close-to-final draft of the manuscript and for her useful comments. UH acknowledges the hospitality and support of the CERN theory division. FK is supported by the German Science Foundation (DFG) under the Collaborative Research Center (SFB) 676 Particles, Strings and the Early Universe.
TMPT is supported in part by NSF grant PHY-1316792 and by the University of California, Irvine through a Chancellor's Fellowship.

\begin{appendix}

\section{EW contributions to mono-jet searches}
\label{app:EWmonoj}

In this appendix, we  discuss the numerical impact of EW contributions of the form $pp \to W \hspace{0.25mm} (\to q \bar q^\prime) + Z^\prime  \hspace{0.25mm} (\to \chi \bar \chi)$ to mono-jet signatures at the LHC, comparing the effects to the corrections associated to the QCD process $pp \to {\rm jets} + Z^\prime  \hspace{0.25mm} (\to \chi \bar \chi)$.  To make this exercise concrete, we consider the recent ATLAS search~\cite{Aad:2015zva}, which is based on $20.3 \, {\rm fb}^{-1}$ of $\sqrt{s} = 8 \, {\rm TeV}$ data. We use {\tt MadGraph5\_aMCNLO} with {\tt NNPDF2.3} PDFs to generate partonic events. The simulated parton-level events were showered with  {\tt  PYTHIA~6} \cite{Sjostrand:2006za} and analysed with the publicly available code {\tt CheckMATE}~\cite{Drees:2013wra}, which relies on {\tt DELPHES 3}~\cite{deFavereau:2013fsa} as a fast detector simulation. To cluster jets we used {\tt FastJet}~\cite{Cacciari:2011ma} employing the anti-$k_t$ algorithm~\cite{Cacciari:2008gp} with  radius parameter $R = 0.4$.

In the ATLAS analysis the following preselection  criteria  are imposed. Events are required to have a reconstructed primary vertex, $E_{T,\rm miss} > 150 \, {\rm GeV}$ and at least one jet with $p_T > 30 \, {\rm GeV}$  and $|\eta | < 4.5$ in the final state.  Events that do not pass certain jet quality requirements or do contain charged leptons or isolated tracks are rejected.  Events having a leading jet with $p_T > 120 \, {\rm GeV}$ and $|\eta| < 2.0$ are selected, if the leading-jet $p_T$ and the $E_{T,\rm miss}$  satisfy $p_T/E_{T,\rm miss} >0.5$. Furthermore, the requirement $\Delta \phi ({\rm jet}, \vec{p}_{T,\rm miss}) > 1.0$  on the azimuthal separation  between the direction of the missing transverse momentum and that of each of the selected jets is imposed. Nine distinct SRs are considered with the following $E_{T,\rm miss}$ thresholds $\{150, 200, 250, 300, 350, 400, 500, 600, 700\} \, {\rm GeV}$. 

\begin{figure}[t]
\begin{center}
\includegraphics[width=0.75\columnwidth]{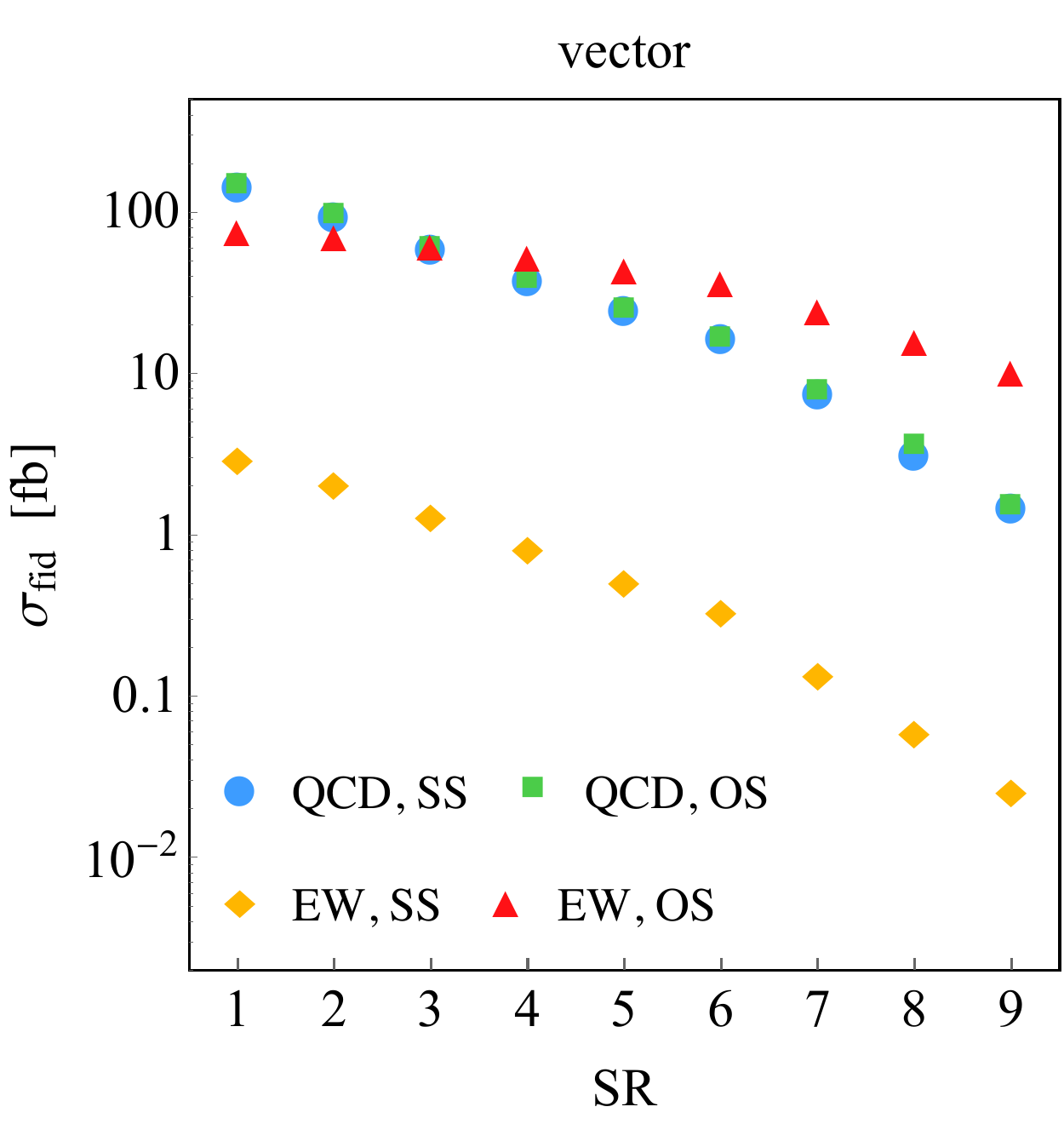} \qquad
\includegraphics[width=0.75\columnwidth]{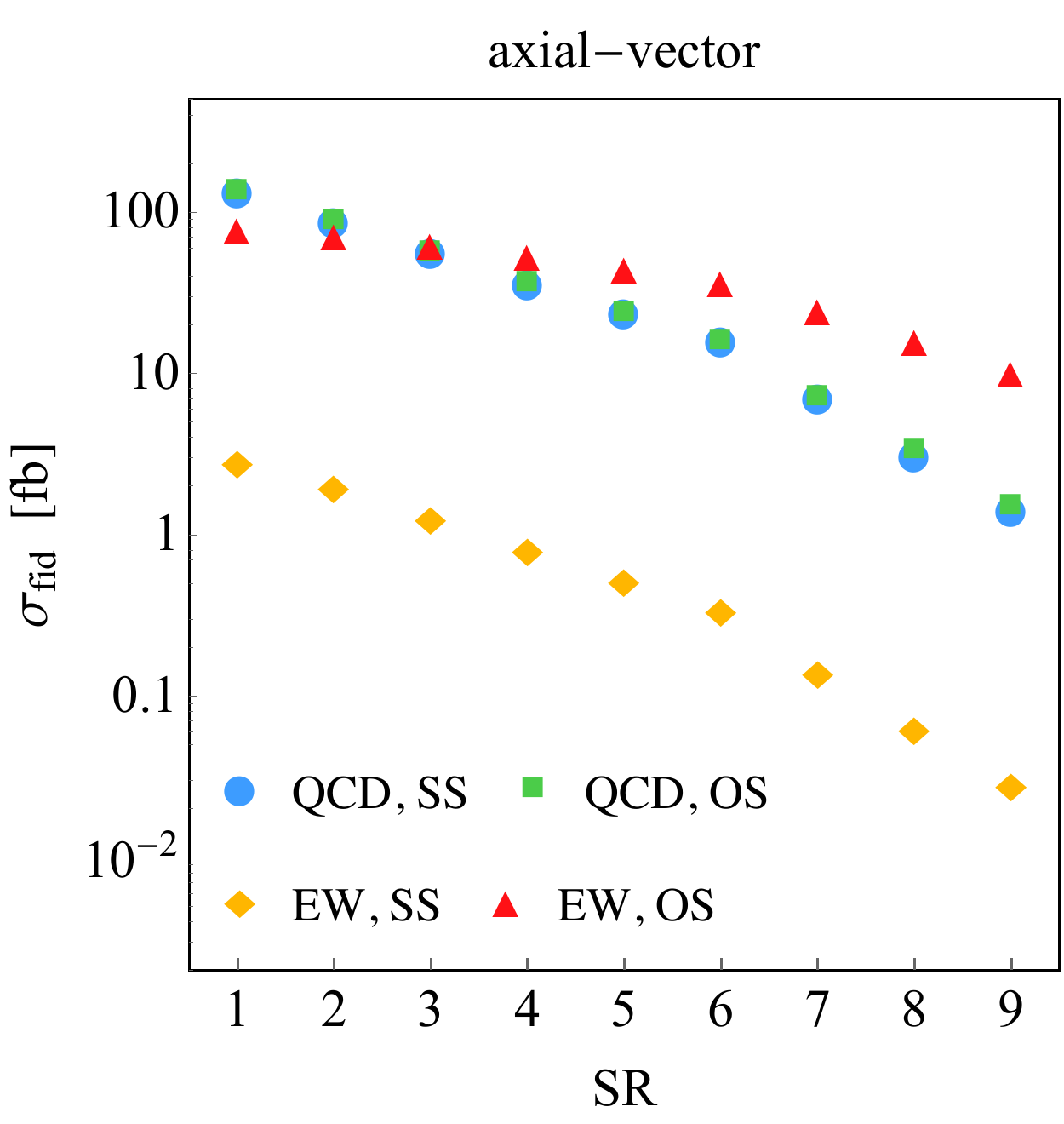}
\end{center}
\vspace{-4mm}
\caption{\label{fig:search} Fiducial mono-jet cross sections for  the SR1 to SR9 selections. The top (bottom) panel shows the predictions for a spin-$1$ simplified  model with only vector (axial-vector) couplings. See text for further explanations.}
\end{figure}

Our results for the fiducial mono-jet cross section corresponding to  SR1 to SR9 are shown in the two panels of Figure~\ref{fig:search}. The plots are based on $M_{Z^\prime} = 1 \, {\rm TeV}$, $m_{\rm DM} = 10 \, {\rm GeV}$ and  all leptonic couplings are taken to be zero. In the left (right) panel we  present the case of pure vector couplings $g_{\rm DM}^V = 1$ and $|g^V_u| = |g^V_d| = 0.25$  (pure axial-vector couplings $g_{\rm DM}^A = 1$ and $|g^A_u| = |g^A_d| = 0.25$). The corresponding total decay width of the $Z^\prime$ boson are $\Gamma_{Z^\prime}= 56.5 \, {\rm GeV}$ and $\Gamma_{Z^\prime}= 55.5 \, {\rm GeV}$ assuming a minimal width. One first observes that the obtained results are to good approximation independent of whether vector or axial-vector exchange is considered. Second, while the fiducial cross sections corresponding to the QCD contributions (blue and green markers) are within statistical uncertainties identical for the SS and OS choices, in the EW case the OS signal strengths (red markers) are, depending on the considered SR, larger than the SS predicitons (yellow markers) by a factor of around~25 to 400. One also sees that, as a result of the constructive interference in the OS case,  starting with SR4 the  EW fiducial cross sections  $\sigma_{\rm fid}^{\rm EW, OS}$ surpass  the corresponding QCD predictions. For the SR7 selection corresponding to $E_{T,\rm miss} > 500 \, {\rm GeV}$, we find for instance $\sigma_{\rm fid}^{\rm EW, OS}/\sigma_{\rm fid}^{\rm QCD, OS} \simeq 330\%$ ($\sigma_{\rm fid}^{\rm EW, OS}/\sigma_{\rm fid}^{\rm QCD, OS} \simeq 350\%$) in the vector (axial-vector) case.  The ratio of the SS fiducial cross sections in SR7 amount instead to $\sigma_{\rm fid}^{\rm EW, SS}/\sigma_{\rm fid}^{\rm QCD, SS} \simeq 2\%$, independently of the choice of mediator. This example nicely illustrates that the mono-$W$ problem is not only relevant for searches that look for a $W$ boson and large $E_{T, \rm miss}$, but would in general also impact mono-jet searches, if it is not resolved.  

\end{appendix}

\end{document}